\documentclass[lettersize,journal]{IEEEtran}
\usepackage{amsmath,amsfonts}
\usepackage{algorithmic}
\usepackage{algorithm}
\usepackage{array}
\usepackage[caption=false,font=normalsize,labelfont=sf,textfont=sf]{subfig}
\usepackage{textcomp}
\usepackage{stfloats}
\usepackage{url}
\usepackage{verbatim}
\usepackage{graphicx}
\usepackage{cite}
\usepackage{multirow}

\begin{document}

\title{Boosting the Targeted Transferability of Adversarial Examples via Salient Region \& Weighted Feature Drop}

\author{
    Shanjun Xu\textsuperscript{\rm 1},
    Linghui Li\textsuperscript{\rm 1},
    Kaiguo Yuan\textsuperscript{\rm 1},
    Bingyu Li\textsuperscript{\rm 2}\\
    \textsuperscript{\rm 1}Beijing University of Posts and Telecommunications\\
    \textsuperscript{\rm 2}Beihang University
}



\maketitle

\begin{abstract}
Deep neural networks can be vulnerable to adversarially crafted examples, presenting significant risks to practical applications. A prevalent approach for adversarial attacks relies on the transferability of adversarial examples, which are generated from a substitute model and leveraged to attack unknown black-box models. Despite various proposals aimed at improving transferability, the success of these attacks in targeted black-box scenarios is often hindered by the tendency for adversarial examples to overfit to the surrogate models. In this paper, we introduce a novel framework based on Salient region \& Weighted Feature Drop (SWFD) designed to enhance the targeted transferability of adversarial examples. Drawing from the observation that examples with higher transferability exhibit smoother distributions in the deep-layer outputs, we propose the weighted feature drop mechanism to modulate activation values according to weights scaled by norm distribution, effectively addressing the overfitting issue when generating adversarial examples. Additionally, by leveraging salient region within the image to construct auxiliary images, our method enables the adversarial example's features to be transferred to the target category in a model-agnostic manner, thereby enhancing the transferability. Comprehensive experiments confirm that our approach outperforms state-of-the-art methods across diverse configurations. On average, the proposed SWFD raises the attack success rate for normally trained models and robust models by 16.31\% and 7.06\% respectively.
\end{abstract}


\section{Introduction}
\IEEEPARstart{R}{ecent} studies \cite{53,41,52} have demonstrated that Deep Neural Networks (DNNs) are vulnerable to adversarial examples, which can lead to misclassification by introducing subtle perturbations to the input. This susceptibility poses notable risks for DNN applications in critical domains such as autonomous driving \cite{42,49} and face recognition \cite{43,48,56}.

\begin{figure}[ht]
	\includegraphics[width=1.0\columnwidth]{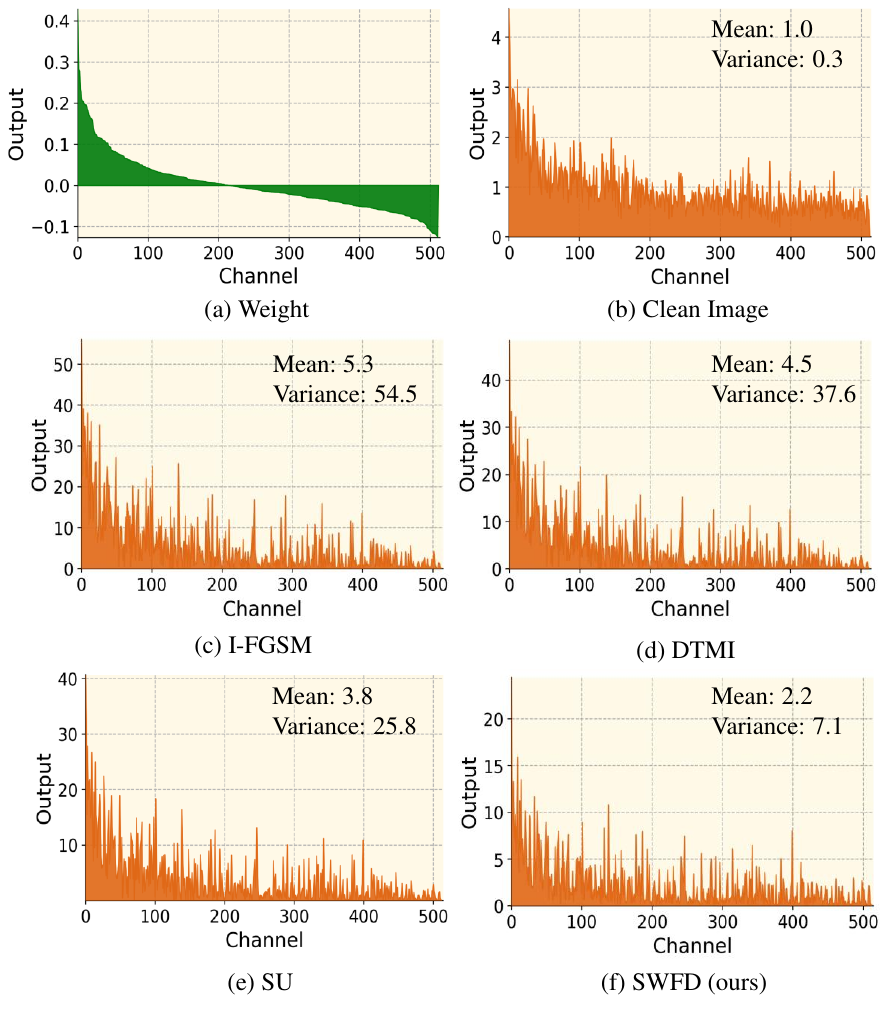}
    \caption{Illustrating the average outputs of the Block-4 in RestNet18. (a) represents the weights of the last linear layer. (b-f) are outputs of clean images and adversarial examples generated by different algorithms. The outputs are plotted according to the indices of the sorted weights.}
	\label{fig:2}
\end{figure}

Numerous approaches \cite{1,3,4,5,7,8} have been proposed for crafting adversarial examples, categorized as white-box and black-box attacks. White-box attacks assume full access to the model's architecture and parameters, whereas black-box attacks are constrained to input-output interactions. Most of adversarial attacks \cite{31,32,33} perform well in white-box scenarios, but they are often less effective in black-box settings, especially against models with defense mechanisms due to the limited knowledge of the target model. A key challenging in black-box attacks is enhancing the transferability of adversarial examples, \emph{i.e.}, their ability to attack the target black-box model from the surrogate model~\cite{55,54}. Thus, in this paper, we concentrates on enhancing the transferability of adversarial examples in targeted black-box scenarios.

Notably, we have observed that when clean images and corresponding adversarial examples are fed into DNNs, targeted adversarial examples crafted by those methods with poor transferability often tend to concentrate on a limited set of features, resulting in overfitting to the surrogate model. As illustrated in Figure \ref{fig:2}, from the perspective of deep layer outputs, the perturbation generation process overly focuses on the surrogate model's target category features, increasing the output variance and reducing the perturbation's generalization across other models (\emph{i.e.}, transferability). In subplots (c-f) of Figure \ref{fig:2}, the mean and variance decrease sequentially (Their mean values are 5.3, 4.5, 3.8, and 2.2, respectively, with variances of 54.5, 37.6, 25.8, and 7.1). Thus, we hypothesize that the output distribution of samples with better transferability are smoother. 

Motivated by this observation, in this paper, we introduce a novel targeted adversarial example attack framework based on Salient region \& Weighted Feature Drop (SWFD) for boosting the transferability. We first propose a weighted feature drop mechanism to prevent the adversarial example generation from becoming overly dependent on a narrow subset of features by diversifying the emphasis across a wider array of features, enhancing the transferability of these adversarial examples. In addition, the underperformance of adversarial examples on different black-box models is likely due to varying decision boundaries. Therefore, we leverage the salient region of the clean image to create auxiliary images, which are employed to further optimize the perturbation, ensuring a robust shift in the feature distribution towards the target category across different models.
Our main contributions are summarized as follows:
\begin{itemize}
\item We propose a novel black-box attack framework aimed at enhancing the transferability of targeted adversarial examples. By smoothing the deep-layer outputs, we improve their adaptability across different models.
\item We design a weighted feature drop mechanism that selects channel-wise features based on weights scaled by norm distribution. This mechanism mitigates the overfitting of adversarial examples to the surrogate model.
\item We leverage the salient region to construct auxiliary images that are used to iteratively optimize the perturbation, effectively aligning the perturbed feature distribution with the target class in a model-agnostic way.
\item The comprehensive experimental results demonstrate that our proposed method has superior transferability, outperforming the state-of-the-art methods in targeted black-box attack scenario.
\end{itemize}


\begin{figure*}[ht]
	\centering
	\includegraphics[width=1.0\linewidth]{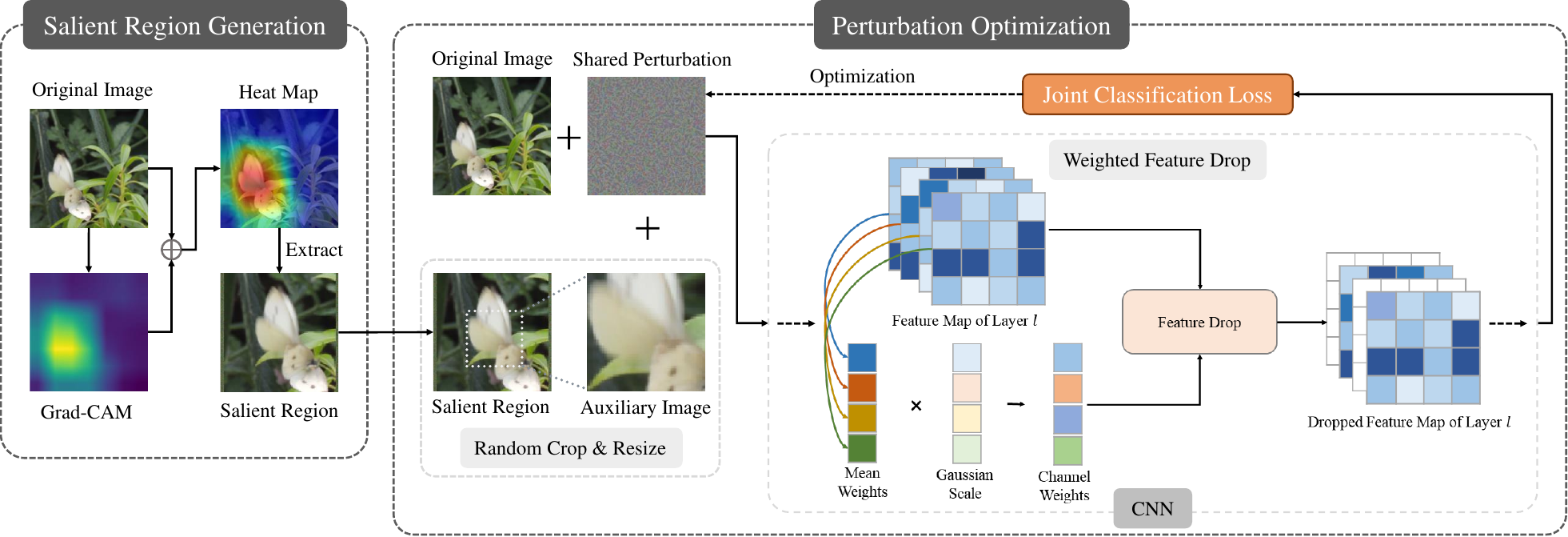}
	\caption{The overview of SWFD framework, which mainly includes two stages: (1) \textbf{Salient region generation.} This stage generates salient regions based on the heatmap; (2) \textbf{Perturbation optimization.} This stage iteratively optimizes the perturbation through the joint classification loss of the original image and auxiliary image based on the weighted feature drop.}
	\label{fig:1}
\end{figure*}

\section{Related Work}
\textbf{Advanced Gradient Calculation.} Inspired by the model training process, several studies have sought to enhance transferability by refining the gradient calculation process. \cite{3} introduce a momentum term to stabilize the perturbation generation process. \cite{4} \cite{4} propose to use the Nesterov to accelerate gradient adaptation in iterative attacks. Translation Invariant (TI) \cite{7} employs a fixed convolutional kernel to smooth gradients, thereby mitigating overfitting. Variance Tuning \cite{8}, on the other hand, utilizes gradients from the vicinity of the data points in the last iteration to alleviate overfitting. Skip Gradient Method (SGM) \cite{9} innovates by modifying the backward propagation path through skip connections. 
Potrip \cite{potrip} Introduce the Poincare distance as the similarity metric to make the magnitude of gradient self-adaptive. PGN \cite{pgn} penalizing gradient norm on the original loss function. These methods create adversarial examples that are heavily reliant on the surrogate model's decision boundary.Therefor, their efficacy can wane when there is a significant divergence between the target and surrogate models.

\textbf{Data Augmentation.} Data augmentation on input images have been demonstrated to significantly enhance the transferability of adversarial examples. Diverse Input (DI) \cite{5} employs random scaling and padding of images. Admix \cite{6} blends input images with those from different categories. Scale-invariant (SI) \cite{4} refines perturbations using scaled versions of images. Self-Universality (SU) \cite{13}, drawing inspiration from Universal Adversarial Perturbation (UAP) \cite{14}, leverages local images to bolster the universality of perturbations. Admix \cite{admix} mix up the images from other categories. SIA \cite{sia} split the image into blocks and apply various transformations to each block. While these methods exhibit transferability, the auxiliary data they utilize may introduce excessive irrelevant information, which can, in turn, limit the extent of transferability achieved.

\textbf{Feature Disruption.} Feature Importance-aware Attack (FIA) \cite{10} and Random Patch Attack (RPA) \cite{46} target significant features that models are likely to consider, fulfilling attack objectives. Attention-guided Transfer Attack (ATA) \cite{11} employs Grad-CAM \cite{12} to obtain feature weights before manipulating them. \cite{47} enhances untargeted attack transferability by integrating random dropout layers, disregarding the importance of various features. While these methods show notable performance in untargeted attacks, their effectiveness is significantly diminished in targeted scenarios. We focus on the targeted black-box attack, from the perspective of output distribution characteristics, proposing a weighted feature drop mechanism designed to smooth the deep-layer output distribution according to scaled weights, thereby achieving higher transferability. Furthermore, we utilize salient regions to refine the perturbation generation, robustly steering the perturbation features towards the target category and mitigating the loss caused by redundant data.


\section{Methodology}
\subsection{Preliminary}
Let $x\in\mathcal{X}\subset\mathcal{R}^{C\times H\times W}$ denote the original image with $C$, $H$, and $W$ representing channels, height, and width of $x$ respectively. The ground-truth label of $x$ is $y_{true}\in\mathcal{Y}=\{1,2,...,M\}$, where $M$ is the category count. We denote $f$ as the white-box surrogate model and $v$ as the black-box model. The target is to generate adversarial example $x_{adv}$ from $f$ to mislead $v$ into classifying it as a different category $y_t\in\mathcal{Y}=\{1,2,...,M\}$, where $v(x_{adv})=y_t$. Adversarail perturbation is constrained by the $L_\infty$ norm: $||x_{adv}-x||_\infty=||\delta||_\infty\le\epsilon$, with $\delta$ as the perturbation and $\epsilon$ as the perturbation limit constant.

The I-FGSM algorithm excels in white-box attacks but struggles with black-box due to the local maxima and overfitting to the surrogate model, reducing the transferability. Methods like DI \cite{5}, TI \cite{7}, MI \cite{3} can mitigate the overfitting in a degree. Following previous work \cite{13}, we adopt the DTMI method, which integrates three methods with I-FGSM, as our benchmark. This targeted attacks can be formulated as:
\begin{gather}
	\label{eq:3}
	g_{i+1}=\mu\cdot g_i+\frac{W\cdot\nabla_\delta J(f(T(x+\delta_i,p)),y_t)}{||W\cdot\nabla_\delta J(f(T(x+\delta_i,p)),y_t)||_1},  \\
	\label{eq:1}
	\delta_{i+1}=\delta_i-\alpha\times sign(g_{i+1}),  \\
	\label{eq:2}
	\delta_{i+1}=Clip_{x,\epsilon}(\delta_{i+1}),
\end{gather}
where $J$ denotes the classification loss, such as cross-entropy (CE). $\delta_i$ and $g_i$ are the adversarial perturbation and gradient at iteration $i$, with $i$ ranging from 0 to $I-1$, where $I$ is the maximum iterations. $T(x+\delta_i,p)$ is used to randomly resizing and padding with probability $p$ as used in the DI method. $W$ is a predefined convolution kernel as used in the TI method. $\mu$ is a decay factor as used in the MI method. $\alpha$ is the iteration step size. The function $Clip_{x,\epsilon}(\cdot)$ ensures that the perturbation $\delta$ meets the $L_\infty$ norm constraint by projecting it near to the original image $x$.


\subsection{Framework Overview}
Figure \ref{fig:1} illustrates the framework of our proposed SWFD, which consists of two main stages. In the salient region generation stage, we extract salient region $x_{sa}$ according to the heatmap generated by the Grad-CAM \cite{12}. In the perturbation optimization stage, we optimize the perturbation $\delta$ by jointly inputting the original image $x$ and the auxiliary image $x_{aux}$ derived from $x_{sa}$ and propagating forward based on the weighted feature drop mechanism in every iteration.

\subsection{Salient Region Generation} \label{sec:3.3}
Deep learning-based image classifiers automatically extract distinguishing features \cite{24}, implying that DNNs possess intrinsic yet weak attention mechanism. \cite{12} \cite{12} suggest that convolutional layers preserve spatial information lost in fully connected layers, offering a balance between high-level semantics and detailed spatial information in the latter layers. Despite various models focusing on different image features, the main features remain consistent across models \cite{11}. To identify the model’s attention areas, we utilize a heatmap aligned with the model’s prediction. Each channel in feature layers acts as a feature extractor. The heatmap is generated using Grad-CAM with the following formula:
\begin{gather}\label{eq:4}
	\alpha^c_k[t]=\frac{1}{Z}\sum_{m}\sum_{n}\frac{\partial f(x)[t]}{\partial A^c_k[m,n]}, \\
	H^t_k=ReLU(\sum_{c}\alpha^c_k[t]\cdot A^c_k),
\end{gather}
where $Z$ is a normalization constant, ensuring attention weights $\alpha_k^c[i]$ fall within the range of $-1$ to $1$. $\alpha_k^c[t]$ denotes attention weights of the $c$-th channel in the $k$-th layer for category $t$. $m$ and $n$ correspond to the height and width of each channel in the $k$-th layer. $H_k^t$ is the heatmap of the $k$-th feature layer for category $t$. Notably, the scale of $H_k^t$ differs from the original image, hence, we bilinearly interpolate the heatmap to match the input resolution for salient region extracting. The ReLU activation function is used to focus solely on positive regions of the heatmap, which are indicative of a positive correlation with the prediction for category $t$. We set $t$ to $y_{true}$ during the heatmap computation.

Once we have the heatmap resized to match the orginal image's scale, we proceed to extract the salient region. Specifically, retain the areas of the heatmap exceeding a predefined threshold $\epsilon_{b}$ (\emph{e.g.}, $\epsilon_b=0.5$). The salient region is then determined by scaling and padding these significant areas to match the size of the original image. 


\subsection{Weighted Feature Drop} \label{sec:3.4}
Overfitting can occur when the perturbation generation overly emphasizes specific features, evident in the rough outputs of DNN deep layers (see Figure~\ref{fig:2}). To address this, we design the weighted feature drop (WFD) mechanism. WFD leverages the principle that adversarial examples with greater transferability typically yield smoother deep-layer. Thus, enhancing transferability involves further smoothing these outputs, which is achieved by spreading focus across a wider feature set. Essentially, features with higher values are more likely to be dropped, alleviating the overfitting.


Let $f_l(x)$ denote the output at layer $l$ for input $x$ in model $f$, with $f_l(x)\in R^{C_l \times H_l \times W_l}$. $C_l$, $H_l$, and $W_l$ represent the number of channels, height, and width of the layer $l$ output data in DNNs, respectively. We calculate the mean weight and channel weight of $i$-th channel at layer $l$ as follows:
\begin{align}
	\label{eq:7}
	Mean^i_l(x)=\frac{1}{H_l\times W_l}\sum_{j=1}^{H_l}\sum_{k=1}^{W_l}f_l(x)[i,j,k],  \\
	\label{eq:8}
	W^i_l(x)=r^i_l\times |Mean^i_l(x)|,\quad r_l\sim N(1,\sigma^2),
\end{align}
where $Mean_l^i(x)$ and $W^i_l(x)$ denote the mean and the channel weight of the $i$-th channel at layer $l$, respectively. $H_l$ and $W_l$ are the height and width of channels at layer $l$. $f_l(x)[i,j,k]$ is the specific value in the $i$-th channel of layer $l$'s output. $r_l = \{r_l^1, r_l^2, \dots, r_l^{C_l}\}$, and $N(1,\sigma^2)$ is a normal distribution with mean $1$ and standard deviation $\sigma$, where $\sigma$ adjusts the channel drop probability. In Equation~(\ref{eq:8}), we apply the absolute value of each channel to prevent negative numbers from affecting the results.

After determining the channel weights at layer $l$, we selectively retain those with lower weights and intentionally drop those with higher weights:
\begin{gather}\label{eq:9}
	f^i_l(x)=\left\{\begin{matrix} 
		b^i_l\cdot f^i_l(x),\quad &\mathrm{if}\enspace W^i_l\le kthvalue(W_l(x),k),  \\  
		0, \quad &\mathrm{otherwise},
	\end{matrix}\right.  \\
	\notag
	k=\left \lfloor p_w\times ||W_l(x)||_0\right \rfloor+C_l-||W_l(x)||_0,  \\
	\notag
	b_l\sim Bernoulli(p_{rnd}),
\end{gather}
where $kthvalue(W_l(x), k)$ represents the $k$-th smallest value in $W_l(x)$.  $\lfloor \cdot \rfloor$ denotes the floor function, and $||\cdot||_0$ denotes the $L_0$ norm, indicating the count of non-zero elements. During this step, we eliminate all channels with weights exceeding $kthvalue(W_l(x), k)$. We opt for $||W_l(x)||_0$ over $C_l$ because dropping channels with a mean of $0$ leaves the outcome unaffected. $b_l = \{b_l^1, b_l^2, \dots, b_l^{C_l} \}$, and $p_{rnd}$ control the probability that $b_l^i$ is equal to 1. The set $b_l$ is utilized to regulate the proportion of channels retained.


\subsection{Perturbation Optimization}

In the perturbation optimization stage, we perform data augmentation by randomly cropping and resizing the salient image $x_{sa}$ to create an auxiliary image $x_{aux}$, ensuring it matches the size of $x_{sa}$. This process generates diverse patterns from $x_{sa}$. Following the previous work \cite{13}, we use the parameter $s = \{s_l, s_{int}\}$ to define the cropping range, where the cropped region encompasses $[s_l,s_l+s_{int}]$ of the original image's area. We denote the random crop and resize operation on $x_{sa}$ as $RCR(x_{sa}, s)$. After applying the WFD mechanism on the surrogate model $f$, we formulate the loss function as follows:
\begin{equation}\label{eq:10}
	L(\delta)=J(f(T(x+\delta)),y_t)+J(f(T(RCR(x_{sa},s)+\delta)),y_t),
\end{equation}

Our objective function is:
\begin{equation}\label{eq:11}
	\arg \min_\delta L(\delta),\enspace s.t.||\delta||_\infty \le\epsilon.
\end{equation}
We use the loss function $L(\delta)$ to guide the optimization of the adversarial perturbation $\delta$. Each iteration can be formulated as:
\begin{gather}\label{eq:12}
	g_{i+1}=\nabla_{\delta_i}L(\delta_i),  \\
	\label{eq:13}
	g_{i+1}=\mu\cdot g_i+\frac{W\cdot g_{i+1}}{||W\cdot g_{i+1}||_1},  \\
	\label{eq:14}
	\delta_{i+1}=\delta_i-\alpha \times sign(g_{i+1}),  \\
	\label{eq:15}
	\delta_{i+1}=Clip_{x,\epsilon}(\delta_{i+1}).
\end{gather}
Consequently, the whole optimization phase achieves smooth the output of the deep layer, mitigating the issue that $x_{adv}$ concentrating excessively on specific features, thus enhancing the transferability.


\section{Experiment}

\subsection{Experimental Settings}

\textbf{Surrogate and Black-box Models.} We employ four distinct models as substitutes to generate perturbations: ResNet50 \cite{34}, DenseNet121 \cite{35}, VGGNet16~\cite{36}, and Inception-v3 \cite{37}. The black-box models include ResNet50, DenseNet121, VGGNet16, Inception-v3, VGGNet19, mobilenet\_v3\_large \cite{38}, PNASNet \cite{39}, and SENet \cite{40}. Additionally, we evaluate two adversarially trained models, DeepAugment (DA) \cite{32}, and stylized ImageNet models \cite{33}.

\textbf{Dataset.} Our testing is conducted on the ImageNet-compatible dataset, introduced by the NIPS 2017 adversarial attacks and defense competition, comprising $1000$ labeled images for targeted adversarial attacks.

\textbf{Evaluation Metric.} Following the work \cite{13}, we adopt the Top-1 (\%) Targeted Attack Success Rate (TASR) as a metric to measure the efficacy of attacks against a black-box model, \emph{i.e.}, the percentage of adversarial examples successfully classified as the intended target category by the black-box model. Thus, methods yielding a higher TASR are indicative of greater targeted transferability.

\textbf{Hyperparameters.} We set the maximum perturbation $\epsilon$ to $16/255$ and the step size $\alpha$ to $2/255$. The max iterations are set to $500$.

\begin{table}[h]
	\centering
	\begin{tabular}{ccccc}
		\hline
		Model & Layer 1 & Layer 2 & Layer 3 & Layer 4 \\
		\hline
		Res50 & Block-1 & Block-2 & Block-3 & Block-4 \\
		Den121 & Block-1 & Block-2 & Block-3 & Block-4 \\
		VGG16 & ReLU-4 & ReLU-7 & ReLU-10 & ReLU-13 \\
		Inc-v3 & Conv-4a & Mix-5d & Mix-6e & Mix-7c \\
		\hline
	\end{tabular}
     \vspace{10pt}
    \caption{Intermediate layers for Weighted Feature Drop. The depth from the shallow Layer 1 to the deeper Layer 4.}\label{table:1}
\end{table}

\begin{table*}[ht]
	\centering
	\renewcommand\arraystretch{1.1}
	\scalebox{0.83}{
		\begin{tabular}{m{1.2cm}<{\centering} m{2.0cm}<{\centering} m{1.75cm}<{\centering} m{1.75cm}<{\centering} m{1.75cm}<{\centering} m{1.75cm}<{\centering} m{1.75cm}<{\centering} m{1.75cm}<{\centering} m{1.75cm}<{\centering} m{1.75cm}<{\centering}}
			\hline
			Surrogate & Method & Inc-v3 & Dense121 & VGG16 & Res152 & VGG19 & MobileNet & PNASNet & SENet \\ \hline
			\multirow{6}{*}{Res50} & CE & 2.1/2.3/2.3 & 27.3/31.4/33.5 & 42.6/44.2/44.2 & 26.2/32.0/32.2 & 35.9/39.1/39.1 & 4.1/4.9/4.9 & 6.7/8.1/8.4 & 22.9/27.4/27.7  \\ 
			~ & CE-SU & 2.1/4.0/4.2 & 20.2/48.7/52.5 & 27.3/57.8/61.5 & 20.2/50.2/54.1 & 24.9/54.7/59.2 & 5.5/11.5/12.0 & 7.0/18.9/20.5 & 16.0/38.6/41.6  \\ 
			~ & \textbf{CE-SWFD} & 4.8/11.7/\textbf{13.4} & 47.8/78.3/\textbf{81.5} & 55.7/85.5/\textbf{88.3} & 45.8/75.9/\textbf{79.0} & 52.0/80.3/\textbf{82.1} & 10.9/25.6/\textbf{26.6} & 17.8/43.9/\textbf{45.9} & 36.3/67.2/\textbf{69.8}  \\ \cline{2-10}
			~ & Logit & 2.3/3.3/3.3 & 53.1/61.3/61.9 & 65.7/75.8/75.8 & 56.7/64.6/65.1 & 59.9/67.8/68.3 & 10.2/11.9/11.9 & 14.9/21.9/22.8 & 45.8/57.5/57.6  \\ 
			~ & Logit-SU & 3.7/4.3/4.4 & 51.2/66.6/66.6 & 66.3/77.7/78.9 & 53.1/66.7/68.6 & 61.3/72.5/75.2 & 11.9/16.7/16.7 & 16.9/25.6/26.9 & 46.9/61.6/64.4  \\ 
			~ & \textbf{Logit-SWFD} & 5.9/13.4/\textbf{14.0} & 51.7/80.2/\textbf{82.7} & 57.1/82.8/\textbf{83.3} & 50.0/80.9/\textbf{82.4} & 51.4/79.5/\textbf{81.1} & 13.2/28.9/\textbf{32.1} & 21.8/49.6/\textbf{53.6} & 40.4/72.2/\textbf{76.1}  \\ \hline
			~ & ~ & Inc-v3 & Res50 & VGG16 & Res152 & VGG19 & MobileNet & PNASNet & SENet \\ \hline
			\multirow{6}{*}{Dense121} & CE & 1.9/2.1/2.1 & 18.4/18.7/18.7 & 21.4/23.0/23.0 & 8.6/10.2/10.4 & 19.8/21.3/21.3 & 3.0/3.9/3.9 & 4.4/5.3/5.3 & 13.3/15.6/15.6  \\ 
			~ & CE-SU & 2.1/5.4/5.5 & 17.5/44.3/49.0 & 21.0/49.7/52.3 & 10.3/31.3/33.2 & 21.5/49.2/51.2 & 4.3/9.3/9.9 & 5.7/21.8/22.0 & 12.0/34.8/36.1  \\ 
			~ & \textbf{CE-SWFD} & 3.7/5.6/\textbf{5.8} & 33.1/56.8/\textbf{59.3} & 43.1/67.0/\textbf{69.2} & 20.3/37.1/\textbf{39.7} & 37.9/61.6/\textbf{64.8} & 8.1/15.0/\textbf{15.2} & 10.6/24.2/\textbf{26.9} & 25.8/51.1/\textbf{52.0}  \\ \cline{2-10}
			~ & Logit & 2.9/3.2/3.3 & 37.1/44.1/44.1 & 48.7/53.2/53.4 & 24.2/29.5/29.6 & 42.2/47.8/48.1 & 6.8/7.2/7.4 & 12.6/16.6/16.7 & 32.2/38.9/38.9  \\ 
			~ & Logit-SU & 3.3/4.4/4.4 & 41.7/49.8/51.1 & 51.4/61.4/61.4 & 27.2/36.5/37.7 & 46.2/56.3/57.3 & 8.0/10.5/10.5 & 15.0/22.1/22.1 & 33.8/45.7/45.7  \\ 
			~ & \textbf{Logit-SWFD} & 2.7/5.7/\textbf{6.4} & 37.8/59.6/\textbf{61.8} & 47.0/70.1/\textbf{72.3} & 25.1/42.1/\textbf{43.9} & 41.5/65.0/\textbf{66.8} & 7.9/14.6/\textbf{14.6} & 14.2/28.2/\textbf{28.3} & 31.7/53.4/\textbf{55.2}  \\ \hline
			~ & ~ & Inc-v3 & Res50 & Dense121 & Res152 & VGG19 & MobileNet & PNASNet & SENet \\ \hline
			\multirow{6}{*}{VGG16} & CE & 0.2/0.2/0.2 & 0.8/1.3/1.3 & 0.9/0.9/0.9 & 0.2/0.3/0.3 & 21.9/22.4/22.4 & 0.3/0.5/0.5 & 0.1/0.1/0.1 & 1.3/1.6/1.8  \\ 
			~ & CE-SU & 0.2/0.4/0.4 & 2.2/3.2/3.5 & 2.1/3.7/3.7 & 0.9/1.3/1.3 & 31.3/54.3/55.9 & 0.6/1.0/1.0 & 0.6/0.8/0.8 & 3.5/5.8/6.0  \\ 
			~ & \textbf{CE-SWFD} & 0.1/1.2/\textbf{1.2} & 2.0/8.0/\textbf{10.1} & 3.8/9.5/\textbf{10.5} & 0.6/3.3/\textbf{3.5} & 46.5/73.1/\textbf{75.6} & 0.7/2.8/\textbf{3.0} & 0.5/3.0/\textbf{3.1} & 5.9/13.9/\textbf{15.2}  \\ \cline{2-10}
			~ & Logit & 0.5/0.6/0.6 & 9.0/11.9/12.1 & 10.4/14.0/14.0 & 3.2/5.3/5.3 & 68.0/72.9/72.9 & 1.5/1.7/1.7 & 2.1/3.1/3.1 & 16.8/21.1/21.5  \\ 
			~ & Logit-SU & 0.6/0.8/0.8 & 10.0/13.7/13.9 & 11.9/14.7/15.3 & 3.5/5.6/5.6 & 70.4/76.2/76.2 & 2.1/2.6/2.6 & 2.7/3.5/3.6 & 18.4/22.7/22.7  \\ 
			~ & \textbf{Logit-SWFD} & 0.3/1.6/\textbf{1.6} & 7.7/18.9/\textbf{20.3} & 9.4/22.0/\textbf{24.2} & 2.4/8.9/\textbf{9.2} & 64.5/86.0/\textbf{86.8} & 1.6/3.8/\textbf{4.1} & 1.5/6.4/\textbf{7.7} & 12.3/30.6/\textbf{32.8}  \\ \hline
			~ & ~ & Res50 & Dense121 & VGG16 & Res152 & VGG19 & MobileNet & PNASNet & SENet \\ \hline
			\multirow{6}{*}{Inc-v3} & CE & 3.3/5.1/5.4 & 4.5/6.1/6.1 & 4.5/6.1/6.1 & 1.7/3.7/3.7 & 2.1/3.5/3.9 & 1.9/2.3/2.5 & 2.9/5.2/5.4 & 3.7/4.6/4.9  \\ 
			~ & CE-SU & 1.5/4.9/5.4 & 1.9/7.3/8.8 & 0.9/3.8/3.8 & 1.0/4.7/5.5 & 1.0/5.1/6.2 & 1.5/3.8/4.5 & 1.7/7.8/9.0 & 1.1/4.9/5.5  \\ 
			~ & \textbf{CE-SWFD} & 3.0/7.4/\textbf{8.0} & 5.6/11.8/\textbf{12.3} & 2.8/6.8/\textbf{6.8} & 3.1/5.6/\textbf{5.6} & 3.5/6.0/\textbf{6.5} & 2.9/5.2/\textbf{6.4} & 3.9/8.5/\textbf{9.3} & 3.4/6.2/\textbf{6.8}  \\ \cline{2-10}
			~ & Logit & 3.0/4.7/5.2 & 4.2/7.2/7.8 & 1.8/4.1/4.5 & 3.0/4.0/4.0 & 2.4/3.8/4.7 & 2.2/3.0/3.0 & 3.8/5.6/5.6 & 3.2/5.2/5.6  \\ 
			~ & Logit-SU & 3.5/6.1/6.3 & 4.9/8.1/9.6 & 2.5/4.8/5.0 & 2.4/5.0/5.0 & 3.3/5.2/5.9 & 2.1/3.4/3.7 & 4.0/7.4/7.7 & 3.1/6.0/6.1  \\ 
			~ & \textbf{Logit-SWFD} & 3.8/6.8/\textbf{7.5} & 4.7/12.4/\textbf{13.2} & 2.5/6.8/\textbf{6.8} & 2.4/5.5/\textbf{5.5} & 2.7/4.9/\textbf{6.1} & 3.2/4.9/\textbf{5.3} & 3.5/8.5/\textbf{9.0} & 3.2/7.5/\textbf{7.5}  \\
			\hline
	   \end{tabular}
    }
    \vspace{10pt}
    \caption{Average TASR (\%) on normally trained models. The TASR after 100/300/500 iterations are listed respectively. The best results of each group are marked in bold. `DTMI' is omitted (\emph{e.g.}, `CE' represents DTMI-CE method).}	\label{table:2}
\end{table*}

\begin{table*}
	\centering
    \scalebox{1.0}{
        \begin{tabular}{cccccccccc}
            \hline
            Surrogate & Method & Inc-v3 & Dense121 & VGG16 & Res152 & VGG19 & MobileNet & PNASNet & SENet \\
            \hline
            \multirow{7}{*}{Res50} & PoTrip & 4.0 & 25.4 & 8.8 & 39.2 & 8.7 & 6.6 & 9.4 & 21.9 \\
            ~ & Admix & 0.2 & 3.9 & 1.1 & 8.5 & 1.1 & 0.6 & 0.2 & 1.3 \\
            ~ & SIA & 1.1 & 18.8 & 12.0 & 35.9 & 11.9 & 10.3 & 3.4 & 16.6 \\
            ~ & PGN & 1.3 & 5.1 & 2.6 & 6.7 & 3.1 & 2.8 & 1.8 & 1.9 \\
            ~ & BPA & 2.2 & 18.3 & 12.3 & 10.2 & 11.4 & 4.5 & 3.5 & 5.0 \\
            ~ & SU & 4.0 & 48.7 & 57.8 & 50.2 & 54.7 & 11.5 & 18.9 & 38.6 \\
            ~ & \textbf{SWFD} & \textbf{11.7} & \textbf{78.3} & \textbf{85.5} & \textbf{75.9} & \textbf{80.3} & \textbf{25.6} & \textbf{43.9} & \textbf{67.2} \\
            \hline
             &  & Inc-v3 & Res50 & VGG16 & Res152 & VGG19 & MobileNet & PNASNet & SENet \\
            \hline
            \multirow{6}{*}{Dense121} & PoTrip & 4.7 & 10.3 & 3.4 & 8.7 & 2.9 & 3.5 & 8.0 & 8.0 \\
            ~ & Admix & 1.9 & 4.5 & 3.8 & 3.3 & 5.8 & 2.5 & 2.4 & 2.8 \\
            ~ & SIA & \textbf{8.3} & 28.5 & 32.7 & 22.8 & 33.8 & \textbf{24.2} & 17.9 & 23.1 \\
            ~ & PGN & 2.2 & 5.0 & 4.3 & 3.4 & 6.0 & 3.0 & 3.2 & 3.4 \\
            ~ & SU & 5.4 & 44.3 & 49.7 & 31.3 & 49.2 & 9.3 & 21.8 & 34.8 \\
            ~ & \textbf{SWFD} & 5.6 & \textbf{56.8} & \textbf{67.0} & \textbf{37.1} & \textbf{61.6} & 15.0 & \textbf{24.2} & \textbf{51.1} \\
            \hline
             &  & Inc-v3 & Res50 & Dense121 & Res152 & VGG19 & MobileNet & PNASNet & SENet \\
            \hline
            \multirow{6}{*}{VGG16} & PoTrip & 0.2 & 0.6 & 0.9 & 0.1 & 10.1 & 0.4 & 0.2 & 1.3 \\
            ~ & Admix & 0.9 & 1.0 & 4.4 & 0.6 & 24.1 & 1.5 & 0.3 & 1.5 \\
            ~ & SIA & 0.9 & 8.0 & \textbf{15.6} & 3.1 & 71.4 & \textbf{9.8} & 2.9 & 13.2 \\
            ~ & PGN & 0.3 & 0.3 & 1.5 & 0.3 & 15.0 & 0.9 & 0.5 & 0.2 \\
            ~ & SU & 0.4 & 3.2 & 3.7 & 1.3 & 54.3 & 1.0 & 0.8 & 5.8 \\
            ~ & \textbf{SWFD} & \textbf{1.2} & \textbf{8.0} & 9.5 & \textbf{3.3} & \textbf{73.1} & 2.8 & \textbf{3.0} & \textbf{13.9} \\
            \hline
             &  & Res50 & Dense121 & VGG16 & Res152 & VGG19 & MobileNet & PNASNet & SENet \\
            \hline
            \multirow{6}{*}{Inc-v3} & PoTrip & 0.6 & 1.7 & 0.6 & 0.4 & 0.5 & 1.2 & 2.5 & 0.4 \\
            ~ & Admix & 0.1 & 0.2 & 0.0 & 0.1 & 0.1 & 0.1 & 0.1 & 0.1 \\
            ~ & SIA & 1.2 & 3.0 & 1.6 & 1.2 & 1.0 & 2.6 & 2.0 & 1.7 \\
            ~ & PGN & 0.1 & 0.4 & 0.1 & 0.0 & 0.2 & 0.0 & 0.4 & 0.1 \\
            ~ & SU & 4.9 & 7.3 & 3.8 & 4.7 & 5.1 & 3.8 & \textbf{7.8} & 4.9 \\
            ~ & \textbf{SWFD} & \textbf{7.4} & \textbf{11.8} & \textbf{6.8} & \textbf{5.6} & \textbf{6.0} & \textbf{5.2} & 7.0 & \textbf{6.2} \\
            \hline
        \end{tabular}
    }
    \vspace{10pt}
    \caption{Average TASR (\%) on normally trained models. The TASR after 300 iterations with CE loss are listed respectively. The best results of each group are marked in bold.}\label{table:6}
\end{table*}


\subsection{Performance Comparison} \label{sec:4.2}

Following the work \cite{13}, we employ the DTMI method as our performance baseline. We strategically choose various layers accoding to the partitioning manner shown in Table \ref{table:1} to apply WFD mechanism for different models. Our proposed SWFD is integrated with several baselines: Cross-Entropy loss with DTMI (DTMI-CE) and Logit loss \cite{30} with DTMI (DTMI-Logit). Furthermore, we conduct comparisons with the Self-Universality (SU) method, using its default parameters: $\lambda=10^{-3}$, $s=(0.1,0)$, and selecting the outputs of Layer 3 for feature similarity computation. We also conducted comparisons with the PoTrip \cite{potrip}, Admix \cite{admix}, SIA \cite{sia}, PGN \cite{pgn}, and BPA \cite{bpa} methods, utilizing their default configurations.

In our implementation of the SWFD, we assign the value $p_w=0.7$, $p_{rnd}=0.7$, and $\sigma=1.3$. For the $RCR$ operation, we utilize $s=\{0.2,0\}$. The WFD mechanism is applied to Layer 3 for both ResNet50 and VGGNet16, and to Layer 4 for DenseNet121 and Inception-v3. 

To demonstrate the superiority of our approach, we conduct transferable experiments to individually attack both normally trained models and robust models.

\begin{table*}[ht]
	\centering
    \begin{tabular}{cccccccc}
        \hline
        $\epsilon$ & Surrogate & Method & SIN\_IN & SIN & DA & Inc-v3$_{adv}$ & IncRes-v2$_{ens}$  \\ \hline
        \multirow{12}{*}{16} & \multirow{6}{*}{Res50} & CE & 32.5 & 2.4 & 4.2 & 0.1 & 0  \\ 
        ~ & ~ & CE-SU & 57.6 & 8.6 & 16.7 & 0.2 & 0  \\ 
        ~ & ~ & \textbf{CE-SWFD} & \textbf{80.5} & \textbf{18.3} & \textbf{30.8} & \textbf{0.4} & \textbf{0.2}  \\ \cline{3-8}
        ~ & ~ & Logit & 66.4 & 5.5 & 11.3 & 0 & 0.1  \\ 
        ~ & ~ & Logit-SU & 69.1 & 6.4 & 13.2 & 0.2 & 0.2  \\ 
        ~ & ~ & \textbf{Logit-SWFD} & \textbf{84.2} & \textbf{21} & \textbf{39.5} & \textbf{0.5} & \textbf{0.3}  \\ \cline{2-8}
        ~ & \multirow{6}{*}{Dense121} & CE & 6.2 & 1 & 2.4 & 0 & 0  \\
        ~ & ~ & CE-SU & 26.3 & 5.6 & \textbf{15.2} & \textbf{0.2} & 0.1  \\ 
        ~ & ~ & \textbf{CE-SWFD} & \textbf{28.5} & \textbf{5.8} & 11.8 & \textbf{0.2} & \textbf{0.2}  \\ \cline{3-8}
        ~ & ~ & Logit & 17.7 & 2.2 & 5.4 & 0 & 0  \\ 
        ~ & ~ & Logit-SU & 24.7 & 3.2 & 8.8 & \textbf{0.1} & 0  \\ 
        ~ & ~ & \textbf{Logit-SWFD} & \textbf{30.7} & \textbf{5.4} & \textbf{12.2} & \textbf{0.1} & \textbf{0.2}  \\ \hline
        \multirow{12}{*}{32} & \multirow{6}{*}{Res50} & CE & 59 & 8.8 & 19.5 & 0.4 & 0.2  \\ 
        ~ & ~ & CE-SU & 84.1 & 21 & 47.8 & 0.2 & 0.2  \\ 
        ~ & ~ & \textbf{CE-SWFD} & \textbf{95.1} & \textbf{45.8} & \textbf{71.2} & \textbf{1.2} & \textbf{0.7}  \\ \cline{3-8}
        ~ & ~ & Logit & 85.1 & 15.4 & 40.7 & 0.4 & 0.4  \\ 
        ~ & ~ & Logit-SU & 87.3 & 20.4 & 46.1 & 0.3 & 0.4  \\ 
        ~ & ~ & \textbf{Logit-SWFD} & \textbf{91.7} & \textbf{50.2} & \textbf{75.4} & \textbf{0.9} & \textbf{1.0}  \\ \cline{2-8}
        ~ & \multirow{6}{*}{Dense121} & CE & 17.5 & 3.3 & 12 & 0.1 & 0.1  \\ 
        ~ & ~ & CE-SU & 47.7 & 14.9 & \textbf{48.2} & 0.4 & \textbf{0.5}  \\ 
        ~ & ~ & \textbf{CE-SWFD} & \textbf{61.8} & \textbf{16.4} & 43.4 & \textbf{0.5} & 0.3  \\ \cline{3-8}
        ~ & ~ & Logit & 30.4 & 8 & 25.1 & 0.2 & 0.2  \\ 
        ~ & ~ & Logit-SU & 44.2 & 11.6 & 35.1 & 0.4 & \textbf{0.5}  \\ 
        ~ & ~ & \textbf{Logit-SWFD} & \textbf{60.5} & \textbf{16.8} & \textbf{45.5} & \textbf{0.6} & 0.4  \\ 
        \hline
    \end{tabular}
    \vspace{10pt}
    \caption{Average TASR (\%) on robust models. The best result in each group marked in bold.}	\label{table:3}
\end{table*}

\begin{table*}
	\centering
    \begin{tabular}{cccccccc}
        \hline
        $\epsilon$ & Surrogate & Method & SIN\_IN & SIN & DA & Inc-v3$_{adv}$ & IncRes-v2$_{ens}$ \\
        \hline
        \multirow{13}{*}{16} & \multirow{7}{*}{Res50} & PoTrip & 7.8 & 0.7 & 2.4 & 0.0 & 0.0 \\
        ~ & ~ & Admix & 0.7 & 0.3 & 0.2 & 0.0 & 0.0 \\
        ~ & ~ & SIA & 5.5 & 0.3 & 0.7 & 0.0 & 0.0 \\
        ~ & ~ & PGN & 3.3 & 0.9 & 1.9 & 0.2 & \textbf{0.3} \\
        ~ & ~ & BPA & 28.0 & 1.5 & 2.4 & 0.0 & 0.0 \\
        ~ & ~ & SU & 54.2 & 8.3 & 15.4 & 0.2 & 0.0 \\
        ~ & ~ & \textbf{SWFD} & \textbf{78.5} & \textbf{16.6} & \textbf{28.6} & \textbf{0.2} & 0.0 \\
        \cline{2-8}
        ~ & \multirow{6}{*}{Dense121} & PoTrip & 5.0 & 1.1 & 3.4 & 0.0 & 0.0 \\
        ~ & ~ & Admix & 2.8 & 0.6 & 1.0 & 0.0 & 0.0 \\
        ~ & ~ & SIA & 22.5 & 1.9 & 5.0 & 0.2 & \textbf{0.1} \\
        ~ & ~ & PGN & 4.7 & 1.4 & 1.4 & 0.1 & \textbf{0.1} \\
        ~ & ~ & SU & 24.6 & \textbf{5.1} & \textbf{13.7} & 0.2 & 0.0 \\
        ~ & ~ & \textbf{SWFD} & \textbf{27.2} & 4.5 & 10.9 & \textbf{0.2} & 0.0 \\
        \hline
        \multirow{13}{*}{32} & \multirow{7}{*}{Res50} & PoTrip & 7.1 & 0.9 & 3.2 & 0.0 & 0.0 \\
        ~ & ~ & Admix & 0.9 & 0.2 & 0.1 & 0.0 & 0.0 \\
        ~ & ~ & SIA & 5.8 & 0.4 & 0.5 & 0.0 & 0.0 \\
        ~ & ~ & PGN & 2.5 & 0.7 & 2.2 & 0.1 & 0.1 \\
        ~ & ~ & BPA & 28.2 & 1.5 & 2.4 & 0.0 & 0.0 \\
        ~ & ~ & SU & 82.8 & 20.2 & 45.4 & 0.2 & 0.2 \\
        ~ & ~ & \textbf{SWFD} & \textbf{93.2} & \textbf{42.9} & \textbf{69.1} & \textbf{0.9} & \textbf{0.7} \\
        \cline{2-8}
        ~ & \multirow{6}{*}{Dense121} & PoTrip & 4.9 & 0.8 & 3.6 & 0.2 & 0.0 \\
        ~ & ~ & Admix & 3.0 & 0.5 & 1.4 & 0.0 & 0.0 \\
        ~ & ~ & SIA & 22.4 & 1.6 & 5.1 & 0.1 & 0.0 \\
        ~ & ~ & PGN & 4.8 & 1.4 & 1.7 & 0.1 & 0.1 \\
        ~ & ~ & SU & 45.8 & 14.6 & \textbf{47.1} & \textbf{0.4} & \textbf{0.5} \\
        ~ & ~ & \textbf{SWFD} & \textbf{59.5} & \textbf{15.6} & 40.8 & 0.2 & 0.3 \\
        \hline
    \end{tabular}
    \vspace{10pt}
    \caption{Average TASR (\%) on robust models. The TASR after 300 iterations with CE loss are listed respectively.}\label{table:7}
\end{table*}

\textbf{Normally trained models transferable attack.} We choose a substitute model from the four aforementioned models to generate the adversarial examples for attacking other black-box models. The attack results are shown in Table \ref{table:2} and \ref{table:6}. We detail the attack results at intervals of 100 iterations, up to 500, to more clearly illustrate the comparative effectiveness of various methods.

Firstly, it is evident that our SWFD attack method outperforms the state-of-the-art methods across various substitute models and loss functions. For instance, utilizing ResNet50 as the substitute model, SWFD achieves an average improvement of 21.31\% with the CE loss and 11.30\% with the Logit loss, compared to the SU method.

Secondly, our attack proves more effective when employing ResNet50 and DenseNet121 as surrogate models over using VGGNet16 and Inception-v3. This result is consistent with previous studies \cite{13,20,29,30}. The inclusion of skip connections in ResNet50 and DenseNet121 significantly mitigates the gradient vanishing issue, especially as the iteration number increases. This characteristic enhances the transferability of the adversarial examples.

\begin{figure}[t]
    \centering
    \subfloat[TASR (\%)]{\includegraphics[width=0.48\columnwidth]{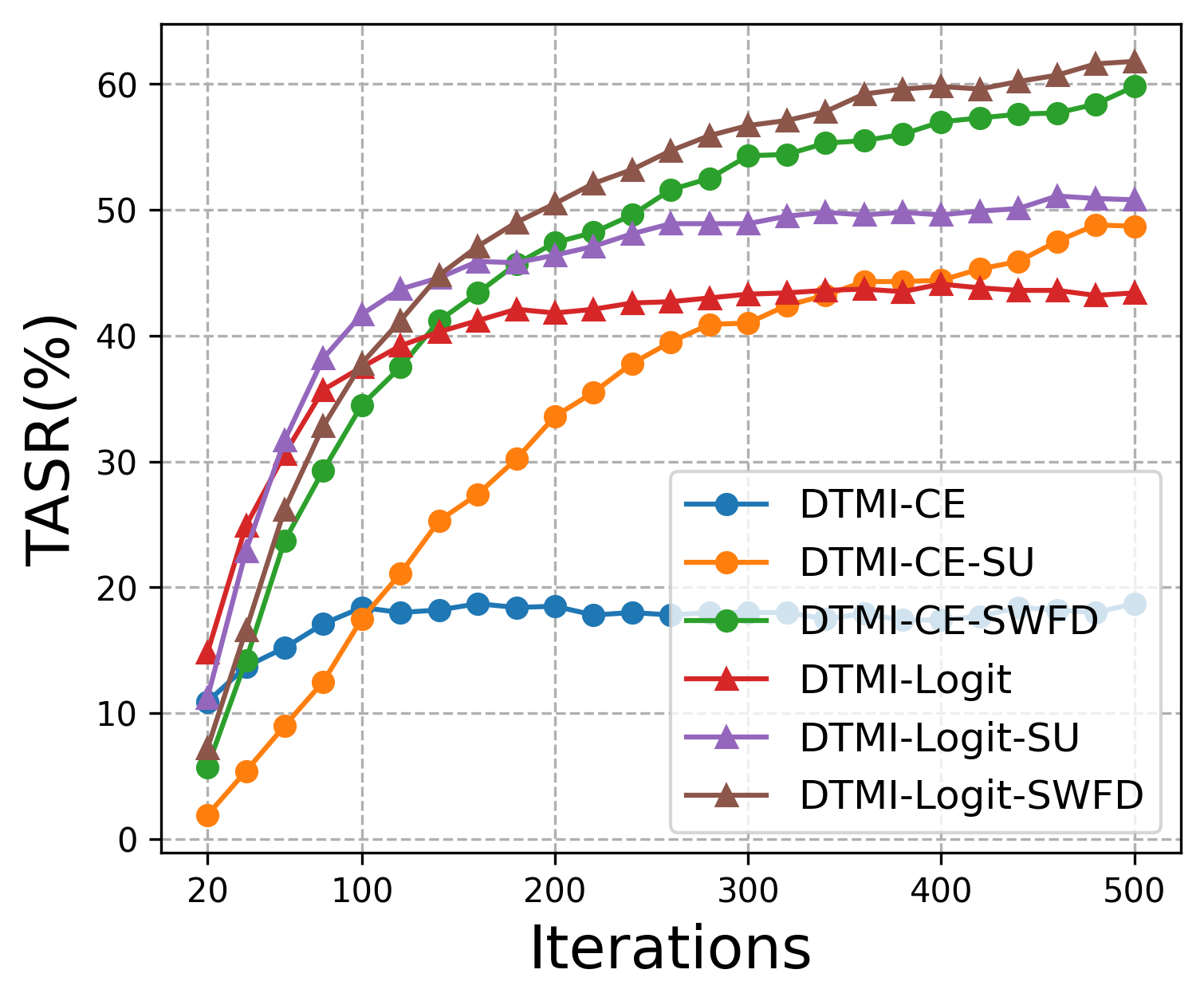}}\label{fig:4.1}
    \subfloat[Gradient Norm]{\includegraphics[width=0.48\columnwidth]{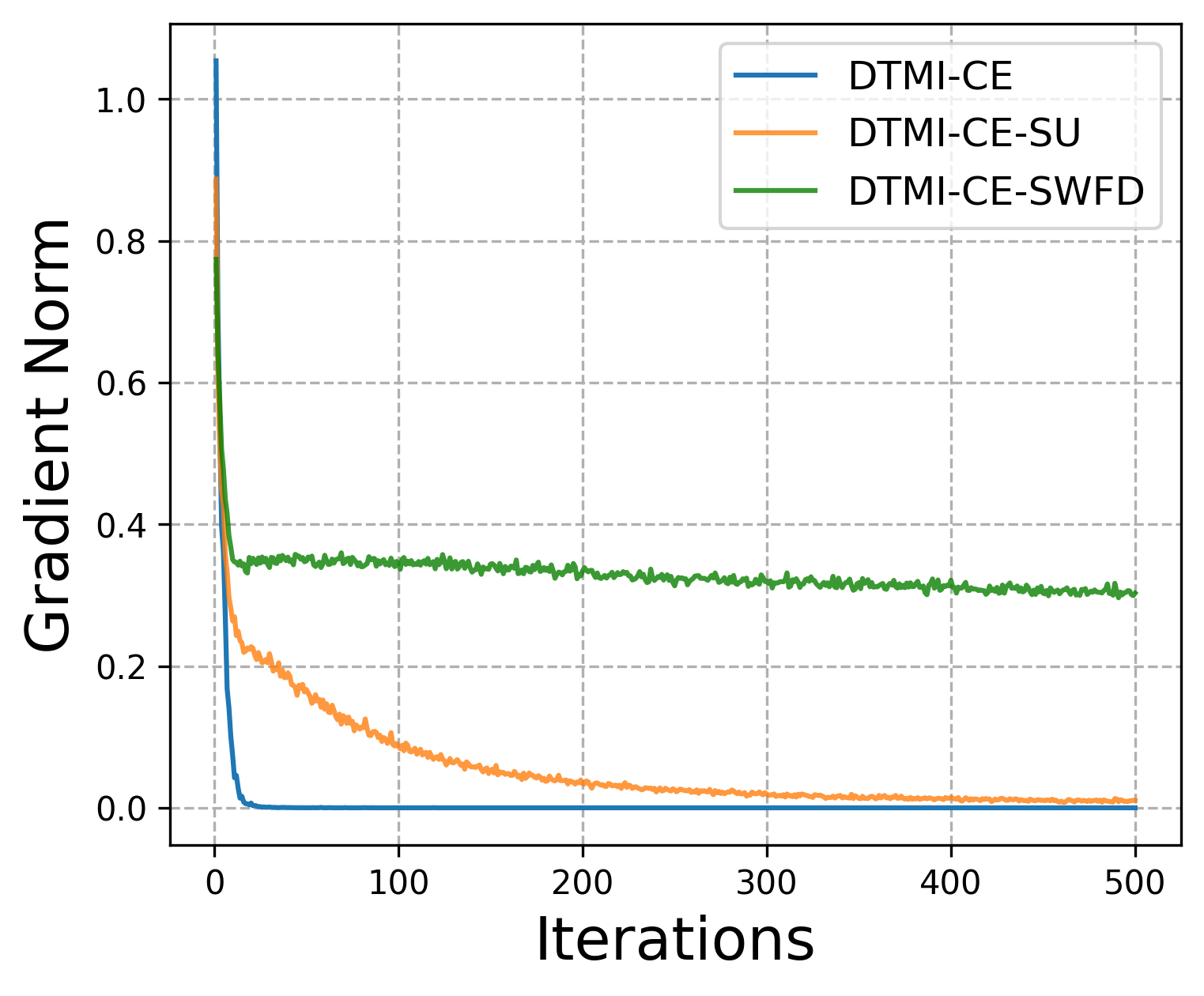}}\label{fig:4.2}
    
    \caption{(a) TASR (\%) when attacking ResNet50 from DenseNet121. (b)~The change in gradient norm as the number of iterations increases.}
    \label{fig:4}
\end{figure}

Thirdly, employing the Logit loss function generally results in better attack performance compared to the CE loss. This is because the CE loss based on the softmax function is susceptible to gradient vanishing during the derivation process, which can diminish its effectiveness. In contrast, our SWFD method effectively mitigates the gradient vanishing issue, as depicted in Figure~\ref{fig:4.2}. This leads to a less significant difference between CE and Logit loss, as illustrated in Figure \ref{fig:4.1}. It highlights the robustness of the SWFD method across different loss functions and its capacity to enhance transferability. Moreover, at lower iteration counts, such as $I=100$, the TASR of the SWFD is slightly lower than that of the SU method when using Logit loss. This is attributed to the SWFD method's dropping of higher-weighted features, promoting a more model-agnostic perturbation update and a slower convergence rate. This phenomenon is observable in Figure~\ref{fig:4.1}, where the DTMI-Logit-SWFD's performance is initially not as strong as DTMI-Logit-SU. However, as iteration increases, while DTMI-Logit-SU reaches convergence at around $300$ iterations, DTMI-Logit-SWFD maintains a steady ascent.

\begin{table}[t]
	\centering
	\begin{tabular}{ccc}
		\hline
		Salient Region & Weighted Feature Drop & TASR (\%)  \\ \hline
		- & - & 23.8  \\ 
		\checkmark & - & 27.3  \\ 
		- & \checkmark & 30.5  \\ 
		\checkmark & \checkmark & \textbf{32.8}  \\
		\hline
	\end{tabular}
    \vspace{10pt}
    \caption{Average TASR (\%) using different parts of our method. TASR is the average result of the four models used as substitute models in turn. `-' indicates that the part is not used, and `\checkmark' indicates it is used.}	\label{table:4}
\end{table}

\textbf{Robust models transferable attack.} 
We evaluate our method against two adversarially trained models, the DeepAugment (DA), and stylized ImageNet models. Utilizing ResNet50 and DenseNet121 as surrogate models, we generate adversarial examples to attack these robust models. The results, presented in Table \ref{table:3}, display the performance after 500 iterations. Notably, our method substantially outperforms the SU method, with improvements of up to 29.3\% using the CE loss and 22.9\% using the Logit loss. These experiments demonstrate that our method not only markedly boosts the effectiveness against normally trained models but also has superior generalization against robust models.


\subsection{Ablation Studies} \label{sec:4.3}

\begin{figure*}[t]
    \centering
    \subfloat[Dense121]{\includegraphics[width=0.25\linewidth]{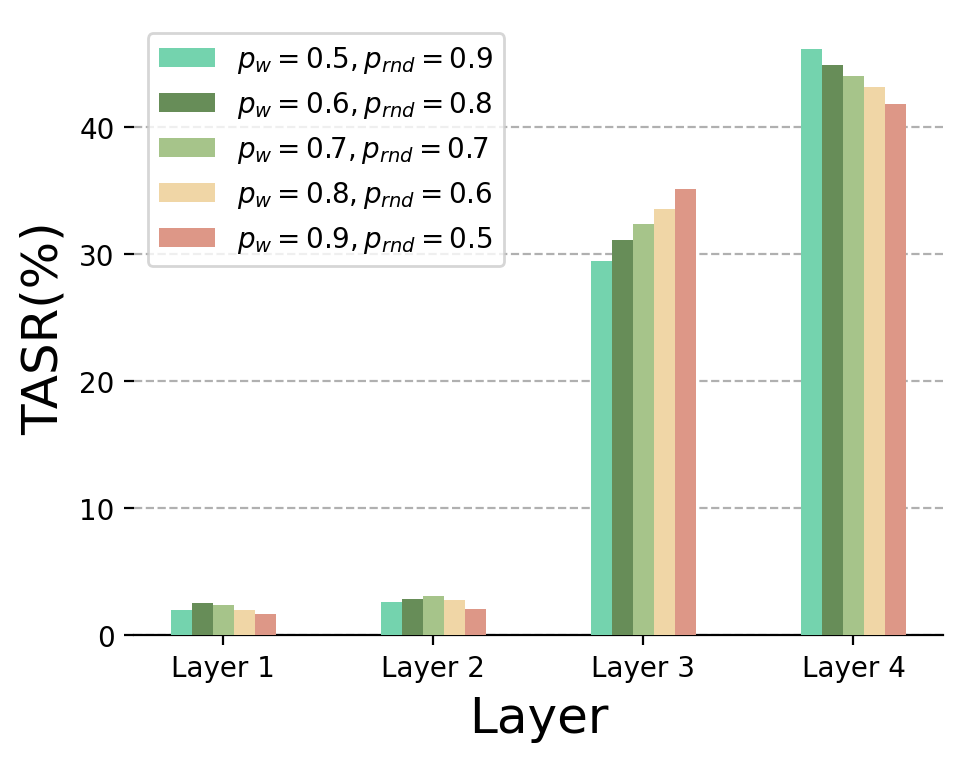}}\label{fig:5.1}
    \hfill
    \subfloat[Inc-v3]{\includegraphics[width=0.24\linewidth]{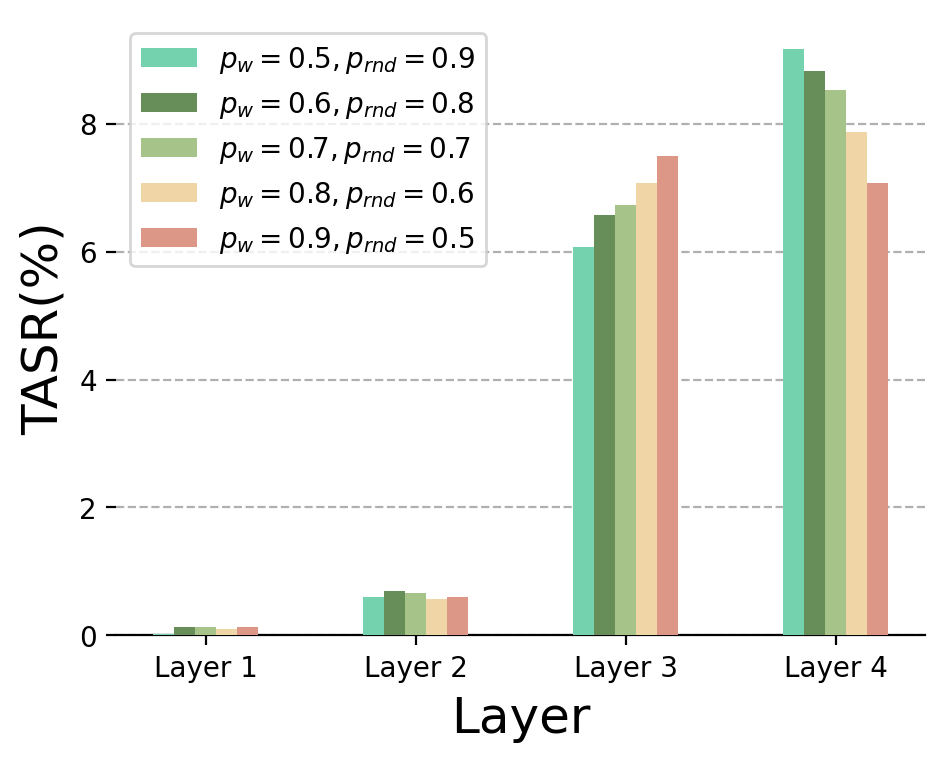}}\label{fig:5.2}
    \hfill
    \subfloat[Res50]{\includegraphics[width=0.25\linewidth]{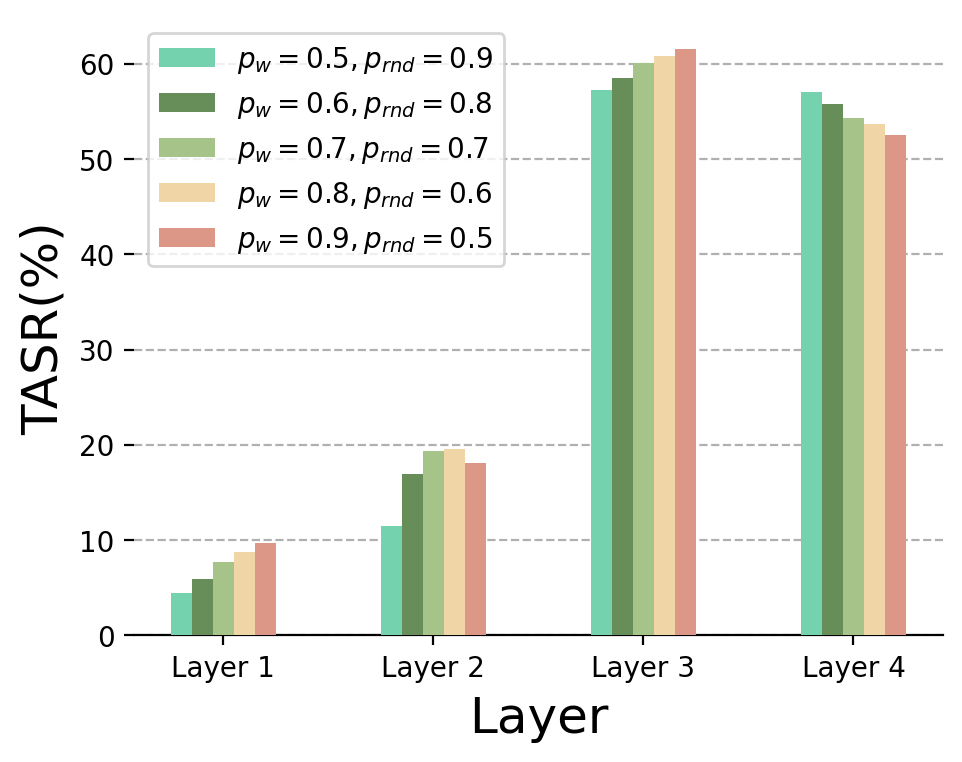}}\label{fig:5.3}
    \hfill
    \subfloat[VGG16]{\includegraphics[width=0.24\linewidth]{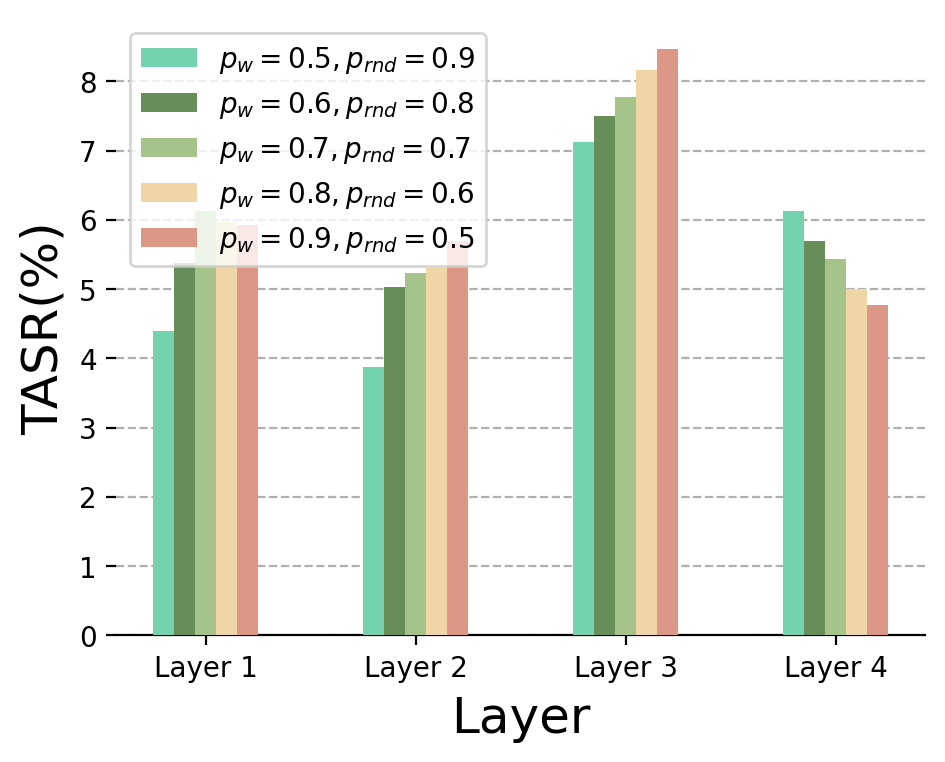}}\label{fig:5.4}
    
    \caption{Average TASR (\%) when applying WFD at different layers with varying $p_w$ and $p_{rnd}$. Sequentially using DenseNet121, Inception-v3, ResNet50, and VGGNet16 as surrogate models. The target black-box models are the remaining three when one is used as the surrogate. The CE loss function is utilized.}
    \label{fig:5}
\end{figure*}

\begin{figure}[t]
    \centering
    \subfloat[Dense121]{\includegraphics[width=0.48\columnwidth]{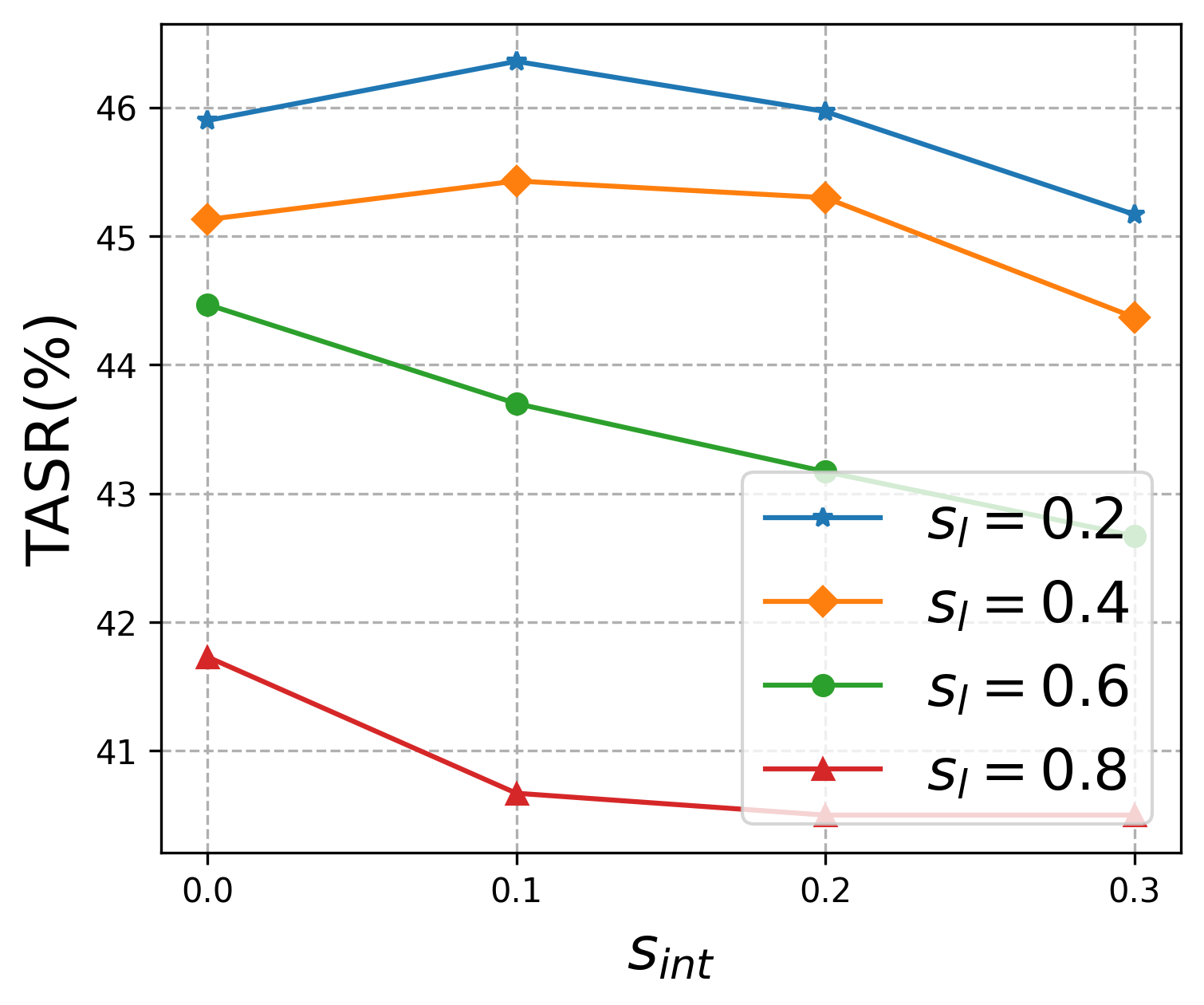}}\label{fig:6.1}
    \subfloat[Inc-v3]{\includegraphics[width=0.48\columnwidth]{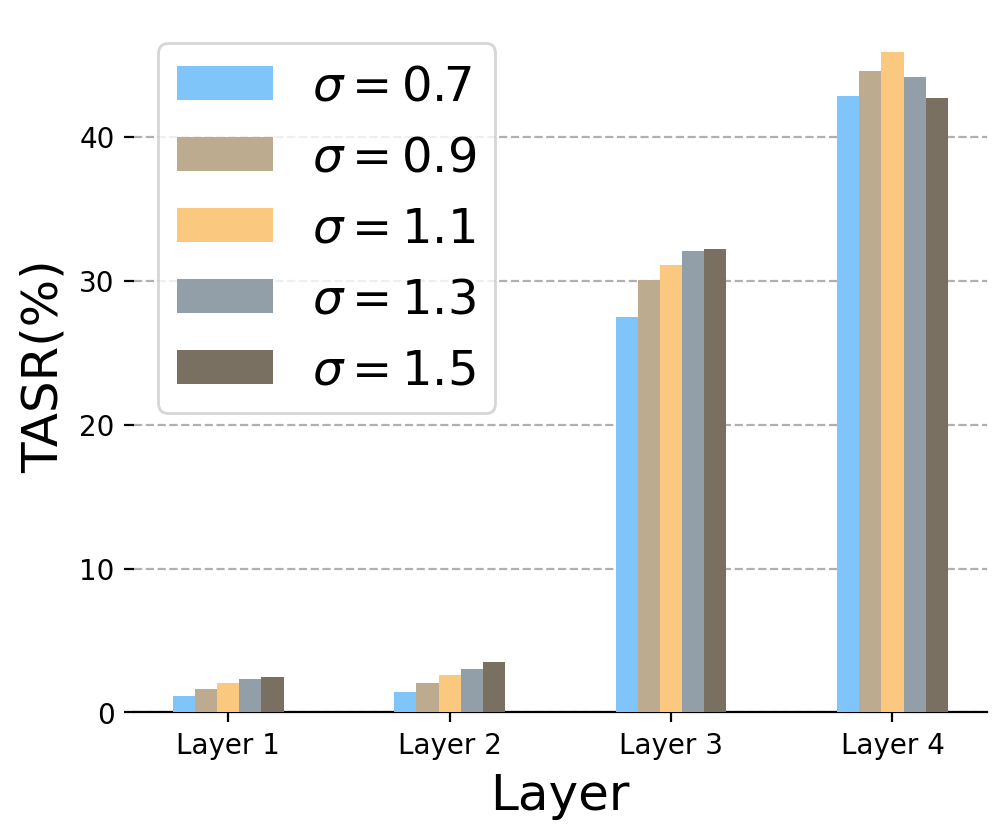}}\label{fig:6.2}
    
    \caption{(a) Average TASR (\%) under different parameters $(s_l, s_{int})$. (b) Average TASR (\%) across different layers $l$ and parameters $\sigma$.}
    \label{fig:6}
\end{figure}

In this section, we investigate the functions of different components and anzlyze the effects of key hyperparameters.

\begin{table}[t]
	\centering
	\begin{tabular}{cccc}
		\hline
		Region & Res50 & VGG16 & Inc-v3  \\ \hline
		Non-salient region & 42.3 & 18.3 & 0  \\ 
		Whole & 62.9 & 20.5 & 0.2  \\ 
		Salient region & \textbf{84.4} & \textbf{43.5} & \textbf{3.3}  \\ 
		\hline
	\end{tabular}
    \vspace{10pt}
    \caption{TASR (\%) with different auxiliary image source regions without WFD. We set $s_l=0.2$, $s_{int}=0.0$.}	\label{table:5}
\end{table}

\textbf{Different components.} We have performed 500 iterations with DenseNet121, Inception-v3, ResNet50, and VGGNet16 as our surrogate models, employing the Logit loss to assess the effectiveness of different components of our proposed SWFD. The experimental results, as shown in Table \ref{table:4}, reveal a significant improvement in the transferability of targeted black-box attack across all components of our model. The weighted feature drop (WFD) mechanism, in particular, has shown a substantial and noticeable enhancement over the Salient Region. It is also noteworthy that the Salient Region, when used in isolation, has already outperformed the 26.4\% achieved by the SU method. These results collectively demonstrate the efficacy of each component of our proposed approach.

\textbf{Hyper-parameters.} Firstly, we assess the average TASR (\%) using various substitute models, exploring different layers and adjusting the parameters $p_w$ and $p_{rnd}$. By setting $s_l=0.2$ and $s_{int}=0.0$, we minimize the randomness in our experiments. As illustrated in Figure~\ref{fig:5}, the WFD mechanism proves more effective for deeper layers, such as the Layer 3 and Layer 4, likely due to their closer relationship with the deep layer. Employing the WFD mechanism in deeper layers facilitates smoother output, achieving greater transferability. Notably, for the Layer 3, parameters with stronger randomness (\emph{i.e.}, $p_w=0.9$ and $p_{rnd}=0.5$) yield superior effects, whereas for the Layer 4, parameters with less randomness and a higher probability of dropping significant features (\emph{i.e.}, $p_w=0.5$ and $p_{rnd}=0.9$) perform better. This disparity may be because, compared to the Layer 3, each channel in the Layer 4 maintains a more potent correlation with each output in the model’s penultimate layer, which has a more direct impact on the final output of the model. Therefore, diminishing the randomness in the Layer 4 proves to be more efficacious. 

Additionally, when using DenseNet121 and Inception-v3 as substitute models, selecting the Layer 4 for the WFD method yields better results than the Layer 3, whereas using ResNet50 and VGGNet16 as substitute models, selecting the Layer 3 yields better results. This could be because DenseNet121 and Inception-v3 better utilize features of Layer 4, while ResNet50 and VGGNet16 find a balance between lower-level features and higher-level semantics in the Layer 3.

Secondly, parameters $(s_l, s_{int})$ determine the area range of the auxiliary image $x_{aux}$, extracted from the salient region of the original image. Unlike the work \cite{13} which directly crops from the original image, our two-step cropping process requires a larger $s_l$ for effective $x_{aux}$. We use DenseNet121 as the surrogate model and Inception-v3, ResNet50, VGGNet16 as target black-box models. The average TASR (\%) is calculated for $s_l\in [0.2,0.4,0.6,0.8]$ and $s_{int}\in[0.0,0.1,0.2,0.3]$, $p_w=0.7$, $p_{rnd}=0.7$, and $\sigma=1.3$. The results, as shown in Figure~\ref{fig:6.1}, reach optimal performance at $s_l=0.2$ and $s_{int}=0.1$. It demonstrate that within a certain range, the smaller the area of $x_{aux}$, the more diverse the patterns that $RCR$ operation can provide, thereby making its data augmentation effect more pronounced.

Thirdly, we test the effect of different $\sigma$, taking $\sigma=[0.7,0.9,1.1,1.3,1.5]$, the larger the $\sigma$, the stronger the randomness, meaning the weight-based drop proportion decreases. The results, as shown in Figure~\ref{fig:6.2}, reveal that parameters with stronger randomness (\emph{i.e.}, $\sigma=1.5$) perform better in shallower layers (\emph{i.e.}, the Layer 1 and 2), while at the Layer 4, $\sigma=1.1$ yields the best results. As we previously analyzed, in the Layer 4, a higher degree of randomness does not necessarily equate to better performance.

\textbf{Different Region.} We validate the effectiveness of the salient region in our method. An image is randomly selected from the ImageNet-compatible dataset. Its non-salient region, whole image, and salient region were used as the source of auxiliary images. The test uses DenseNet121 as the substitute model, Inceptionv3, ResNet50, and VGGNet16 as the target black-box models. To better reflect the differences, we did not use the WFD method. And to reduce randomness, the test is conducted 50 times on 32 copies of the image, totaling 1600 attacks with $I=500$. The results, shown in Table~\ref{table:5}, calculate TASR as $\frac{\mathrm{Number\ of\ Successful\ Attacks}}{1600}\times100\%$. Using the salient region achieve the best performance, demonstrating that the choice of different region significantly impacts the attack effect. Using the whole image is also effective since it contains the salient region, which allows feature transfer robustly. Nevertheless, non-salient regions of the image may disrupt the direction of perturbation updates, leading to worse effects. This underscores the importance of extracting the salient region from the image.


\section{Conclusion}

In this paper, we introduce the Salient Region \& Weighted Feature Drop (SWFD), an innovative method engineered to enhance the targeted transferability of adversarial examples. Our proposed approach significantly alleviates the issue of adversarial examples overfitting to substitute models by employing a selective channel-wise feature drop mechanism. This mechanism eliminates features with larger values according to weights scaled by norm distribution, resulting in a smoother output of the model's deep layer and a reduction in overfitting tendencies. Moreover, the proposed SWFD capitalizes on the concept of salient regions within input images and construct auxiliary images based on the extracted salient regions. By iteratively optimizing the perturbation, the method enables the perturbation feature to be transferred to the target category in a model-agnostic manner. This feature significantly amplified the transferability of adversarial examples. We have rigorously tested our proposed method through a series of comprehensive experiments to verify the effectiveness. The results reveal that compared to state-of-the-art transfer-based attacks, our proposed SWFD method generates adversarial examples with superior transferability. This effectiveness is evident when attacking both normally trained models and those fortified with robustness features.


\bibliographystyle{IEEEtran}
\bibliography{IEEEabrv,mybib}

\end{document}